\documentclass[]{spie}

\usepackage{amsmath,amsfonts,amssymb}
\usepackage{graphicx}
\usepackage[colorlinks=true, allcolors=blue]{hyperref}

\title{The TRILL project: increasing the technological readiness of Laue lenses}

\author[a]{Lisa Ferro}
\author[b]{Enrico Virgilli}
\author[a]{Miguel Fernandes Moita}
\author[a,b]{Filippo Frontera}
\author[a,b,e]{Piero Rosati}
\author[a,b,e]{Cristiano Guidorzi}
\author[c]{Claudio Ferrari}
\author[a]{Riccardo Lolli}
\author[b]{Ezio Caroli}
\author[b]{Natalia Auricchio}
\author[b]{John B. Stephen}
\author[f]{Stefano Del Sordo}
\author[f]{Carmelo Gargano}
\author[e]{Stefano Squerzanti}
\author[d]{Mauro Pucci}
\author[h]{Olivier Limousin}
\author[h]{Aline Meuris}
\author[h]{Philippe Laurent}
\author[h]{Hugo Allaire}
\affil[a]{Department of Physics and Earth Science, University of Ferrara, Via Giuseppe Saragat 1/C, Ferrara (FE), 44122, Italy}
\affil[b]{INAF/OAS of Bologna, Via Piero Gobetti 93/3, Bologna (BO), 40129, Italy}
\affil[c]{Institute of Materials for Electronics and Magnetism (CNR-IMEM), Parco Area delle Scienze 37/A, Parma (PR), 43124, Italy}
\affil[d]{National Institute of Optics (CNR-INO), \\
Largo Enrico Fermi 6, Florence (FI), 50125, Italy}
\affil[e]{INFN of Ferrara,  Via Giuseppe Saragat 1/C, Ferrara (FE), 44122, Italy}
\affil[f]{INAF/IASF of Palermo, Via Ugo La Malfa 153, Palermo (PA), 90146, Italy}
\affil[h]{Université Paris-Saclay, Université Paris Cité, CEA, CNRS, AIM, 91191, Gif-sur-Yvette, France}

\authorinfo{Further author information: (Send correspondence to L. Ferro, E. Virgilli)\\Lisa Ferro: E-mail: lisa.ferro@unife.it\\Enrico Virgilli: E-mail:enrico.virgilli@inaf.it }

\begin{document} 
\maketitle

\begin{abstract}
Hard X-/soft Gamma-ray astronomy ($\mathrm{>}$100 keV) is a crucial field for the study of important astrophysical phenomena such as the 511 keV positron annihilation line in the Galactic center region and its origin, gamma-ray bursts, soft gamma-ray repeaters, nuclear lines from SN explosions and more. However, several key questions in this field require sensitivity and angular resolution that are hardly achievable with present technology. A new generation of instruments suitable to focus hard X-/soft Gamma-rays is necessary to overcome the technological limitations of current direct-viewing telescopes. One solution is using Laue lenses based on Bragg’s diffraction in a transmission configuration. To date, this technology is in an advanced stage of development and further efforts are being made in order to significantly increase its technology readiness level (TRL). To this end, massive production of suitable crystals is required, as well as an improvement of the capability of their alignment. Such a technological improvement could be exploited in stratospheric balloon experiments and, ultimately, in space missions with a telescope of about 20 m focal length, capable of focusing over a broad energy pass-band. We present the latest technological developments of the TRILL (Technological Readiness Increase for Laue Lenses) project, supported by ASI, devoted to the advancement of the technological readiness of Laue lenses. We show the method we developed for preparing suitable bent Germanium and Silicon crystals and the latest advancements in crystals alignment technology. 
\end{abstract}


\section{Laue Lenses: concentrators for Hard X-/Soft Gamma-rays}
Laue lenses are a type of hard X-/Gamma-ray focusing optics based on Bragg's law of diffraction in crystals. In this section we will briefly describe the basic concept of a Laue lens. 
\subsection{Laue lenses basics}
\label{sec:laue_lens}  
\begin{figure}
    \centering
    \includegraphics{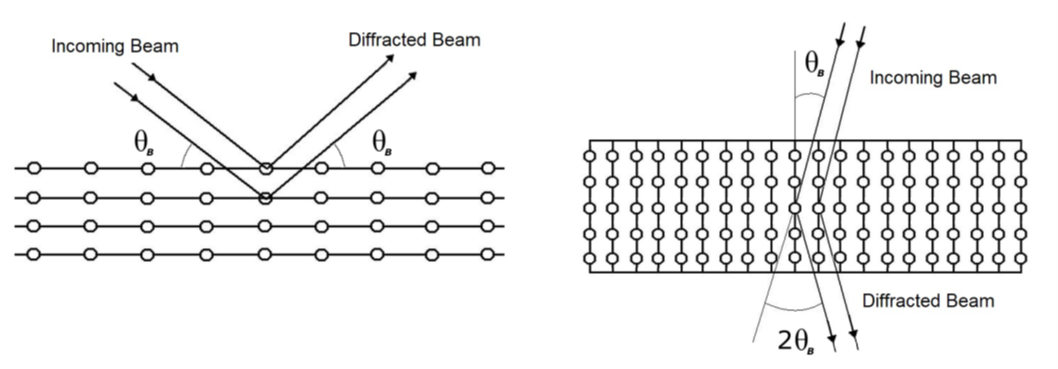}
    \caption{Left: Bragg's diffraction in reflection (Bragg) configuration. The diffracted beam is reflected by the crystal in the same region of space as the incoming beam. Right: Bragg's diffraction in transmission (Laue) configuration, in which the beam crosses the crystals and is diffracted in the opposite region of space with respect to the incoming beam.}
    \label{fig:Bragg_vs_Laue}
\end{figure}

Bragg's law of diffraction in a crystal lattice can be written as\cite{Zachariasen}:
\begin{equation}
    2 d_{hkl}\sin \theta_B = n \frac{hc}{E}
\end{equation}
where $\theta_B$ is the Bragg's angle, i.e. the angle between the chosen diffraction planes of the crystal and the incident beam, $d_{hkl}$ is the inter-planar spacing of the diffracting planes, $(hkl)$ are the Miller indexes of the planes, $E$ is the energy in keV of the gamma-ray photons and $n$ is the order of diffraction.
If we consider a simple, flat crystal, and we imagine that the crystal divides the space in two regions, we can define two possible configurations:
\begin{enumerate}
    \item Bragg configuration, or "reflection", geometry, in which the diffracted beam is in the same region of space as the beam incoming on the crystals (Fig. \ref{fig:Bragg_vs_Laue}, left).
    \item Laue configuration, or "transmission", geometry, in which the diffracted beam crosses the crystal and is transmitted in the region of space opposite to the incoming beam (Fig. \ref{fig:Bragg_vs_Laue}, right). 
\end{enumerate}

Bragg's diffraction in transmission geometry is effective up to energies of a few MeV, meaning that it is possible to build hard X-/Gamma ray optics based on the diffraction process. Specifically, the transmission configuration can be advantageous in an astrophysical context since it allows us to obtain a larger effective area with respect to focusing techniques based on multilayer optics. 

\begin{figure}[t!]
    \centering
    \includegraphics[scale = 0.25]{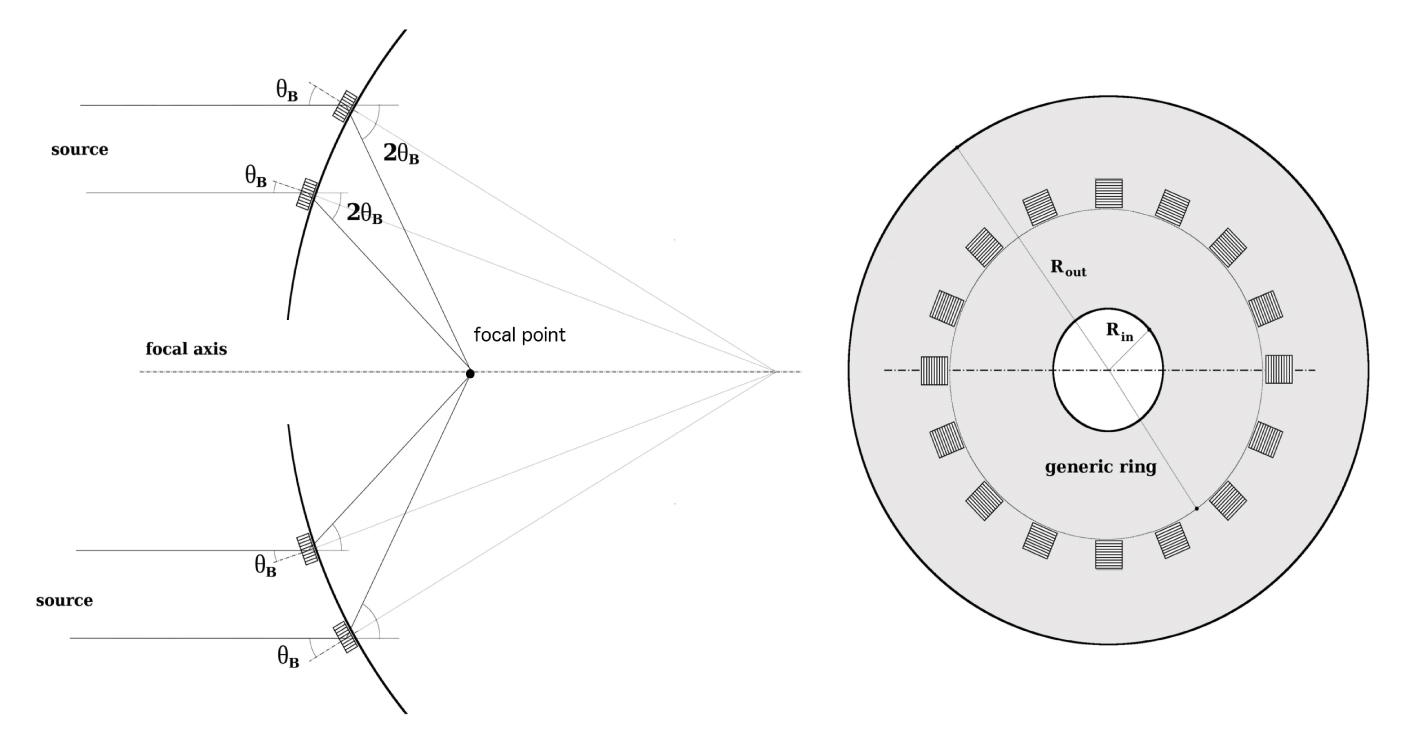}
    \caption{Side and top view of a Laue lens. The crystals are positioned on a spherical support in such a way that the radiation coming from the sky, parallel to the optical axis of the lens, interacts with the crystals and is focused. The diffraction planes of the crystals are oriented in such a way that the angle between them and the incoming X-ray beam is equal to Bragg's angle $\theta_B$, so the angle between the diffracted beam and the incoming beam is $2\theta_B$. The focal length is equal to half the curvature radius of the spherical cap.}
    \label{fig:laue_lens_schematics}
\end{figure}

A hard X-/soft gamma-ray optics based on Bragg's law in transmission configuration is called a Laue lens.
A Laue lens can be visualized as a spherical cap covered by crystal tiles oriented in such a way that the radiation coming from the sky is transmitted through the crystals and focused towards a point which is the lens focus, similarly to optical lenses. At the focus, a suitable position sensitive detector is placed. It can be shown that the focal distance of the lens $f$ is equal to half of the curvature radius of the spherical cap \cite{frontera2011} (Fig. \ref{fig:laue_lens_schematics}) . 

The simplest possible distribution for the crystals on the cap is to place the crystals in concentric rings at a distance $r$ from the axis of the lens. Using the Bragg's law, we can express the centroid of the energy spectrum of the photons diffracted by each ring as\cite{frontera2011}:

\begin{equation}
\label{eq:e_vs_rad}
    E =  \frac{hc}{2d_{hkl}\sin \Big [ \frac{1}{2} \arctan \Big( \frac{r}{f} \Big) \Big]} \sim \frac{hc}{d_{hkl}} \frac{f}{r}
    \end{equation}

which means that hard photons are diffracted by the crystals in the innermost rings of the lens, while soft photons are diffracted by the outermost rings of the Laue lens.
The approximation in Eq.\ref{eq:e_vs_rad} holds true for small Bragg's angles, which is true for hard x-/soft gamma-rays, being of the order of 1 degree or less.
Thanks to the technology of Laue lenses, we will be able to build lightweight ($<$150 kg), stable, broadband optics which will enable real concentration of hard X-/Gamma-rays up to 700 keV and with a Point Spread Function (PSF) with a Half Power Diameter (HPD) of the order of  30 arcsec or less, at 20 m focal distance.

A comparison with one of today's most advanced high energy instruments, the IBIS aboard INTEGRAL, whose angular resolution is about 12 arcmin, shows the quantum leap in imaging capabilities that would be gained by an instrument of this type, which will allow high sensitive spectroscopy and will enable polarimetry in the high energy domain based on Compton kinematics.  Such an improvement in terms of sensitivity and new physical information would allow us to answer some of the most debated questions in high-energy astrophysics, such as the origin of the annihilation line from the Galactic center, the study of GRB afterglow emission, emission spectra of blazars and AGN galaxies, studies on the nuclear lines from supernova (SN) explosions and many more. We refer the reader to the two white papers submitted to the European Space Agency for its long term program "Voyage 2050" for more details on the science goals. \cite{Guidorzi2021,Frontera2021}.

 \subsection{Crystals for Laue lenses}
The fundamental requirements for a Laue lens are its energy pass-band, its effective area in that energy range, its focusing capability and, ultimately, the instrument sensitivity. These requirements can be optimized with a correct selection of crystals having the appropriate material, thickness and diffraction planes. 

 Several types of crystals can be used to build a Laue lens, however for our applications we considered two main families: perfect crystals and mosaic crystals.
 Ideally, a perfect crystal is a crystal in which the lattice shows negligible imperfections of any kind: no dislocations and high chemical purity.

 The rocking curve of a perfect crystal, i.e. the measurement of the intensity of the diffracted beam with respect to the angle at which the crystal is oriented, is a Gaussian centered on the Bragg's angle and its Full Width at Half Maximum (FWHM) is called Darwin width \cite{Zachariasen}. The energy passband of a perfect crystal is very sharp, centered on the values that satisfy Bragg's condition. This means that perfect crystals are very good monochromators, but they are not the best solution if we want to build, as in our case, a broadband instrument with a smooth effective area.

 Instead, a mosaic crystal, is a type of crystal which can be described as a mosaic of microscopic perfect crystals, called {\it crystallites}. In a mosaic crystal, all the crystallites are slightly misaligned around an average direction. The distribution of the crystallites misalignment can be approximated by a Gaussian:
 \begin{equation}
     W (\delta) = \frac{1}{\sqrt{2\pi \eta}}\exp{\bigg(-\frac{\delta^2}{2\eta}\bigg)}\;,
 \end{equation}
 where $\delta$ is the angular deviation of the crystallites with respect to their average direction. The FWHM of the distribution  $\beta = 2.355\,\eta$ is called the crystal {\it mosaicity}.
The presence of a non-zero mosaicity has the effect of increasing the energy passband of the crystals, since it increases the range of angles satisfying the Bragg's condition. The downside is that this wide energy passband impacts on the size of the diffracted image produced by a mosaic crystal. Due to the mosaic spread, the diffracted photons are focused over an enlarged area, whose extension also depends on the focal length. This effect is called {\it mosaic defocusing} and it slightly enlarges the Point Spread Function (PSF) of a mosaic crystal with respect to one of a perfect crystal. 

\begin{figure}[t!]
    \centering
    \includegraphics[scale = 0.35]{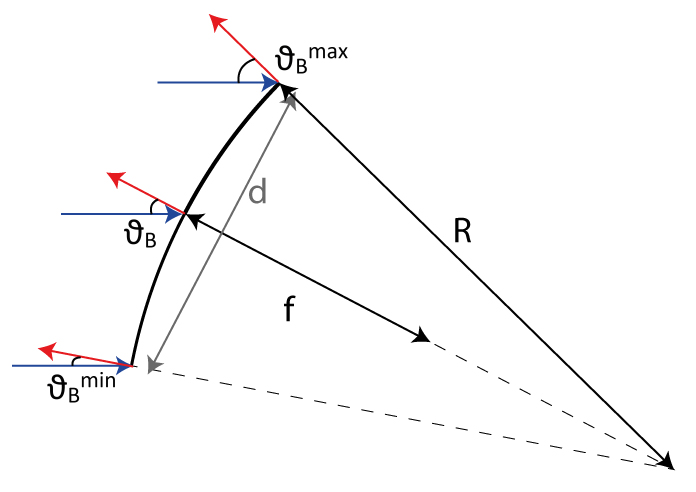}
    \caption{Explanation of the concentration effect given by cylindrical bent crystal: due to the fact that the crystal is bent, the average direction of the diffraction planes (red arrows) varies continuously inside the crystals, which means that the Bragg's angle of the incoming radiation (blue arrows) changes accordingly. The result is that the X-ray beam is concentrated in a point at a distance $f = R/2$ and that the crystal's energy pass-band is enlarged of a quantity which depends on the length of the focusing direction $d$ of the crystal, its curvature radius $R$ and its average diffracted energy.} 
    \label{fig:curvature}
\end{figure}

With our past HAXTEL project \cite{frontera2008}, flat mosaic crystals of copper were successfully used to build a prototype of a Laue lens with a low ($<$6 m) focal length. However, flat crystals (perfect or mosaic) show important limits: first, they do not possess intrinsic concentration capability: 
the size of the resulting image is comparable to the cross section of the crystal itself. 
Second, the diffraction efficiency for a flat crystal cannot be higher than 50 \% \cite{Zachariasen}. Instead, with {\it bent crystals} we can overcome these limitations. 
{While in bent mosaic crystals the average direction of the planes varies according to a curvature radius, in bent perfect crystals the diffraction planes themselves are bent. 
Indeed, for some specific crystallographic configurations a secondary 
curvature is induced by impressing a primary curvature to a perfect 
crystal. This secondary or internal curvature results in an increased 
angular spread of the planes called {\it quasi-mosaicity} \cite{Authier98,Camattari2015}.}
The easiest type of curvature which can be obtained is a cylindrical curvature. A crystal, either perfect or mosaic, bent in this way acts as a concentrator along the bent direction, focusing the radiation coming from an astrophysical source to a focal distance equal to half the curvature radius of the crystal itself\cite{virgilli2014} (Fig. \ref{fig:curvature}).
Thus, through bent crystals it is then possible to decrease the size of the PSF and to enlarge the energy passband, since a wider range of Bragg's angle can be accepted by a bent crystal.
In addition, depending on the arranged internal structure of the atoms within the crystal, their diffraction efficiency can be higher than 50\%.

\begin{figure}[t!]
    \centering
    \includegraphics[scale = 0.8]{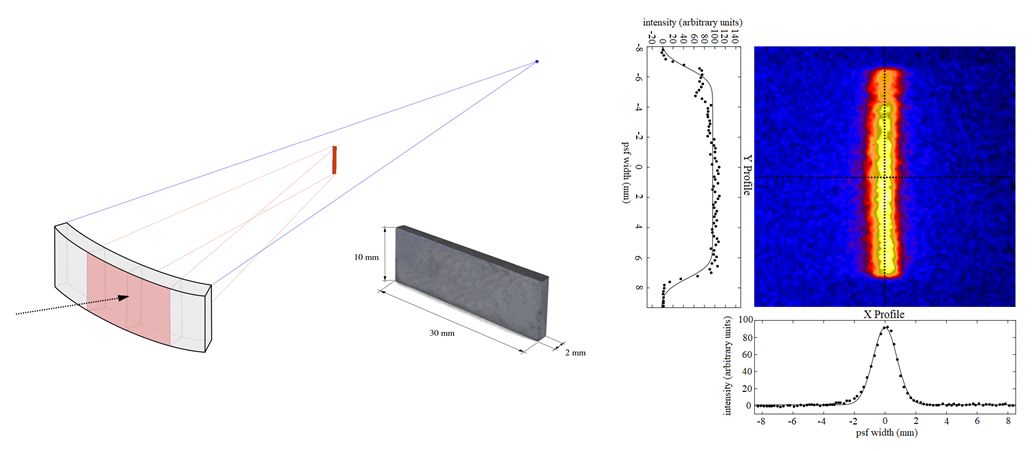}
    \caption{Right: Schematics of the concentration effect of one cylindrical bent, 30$\times$10 mm$^2$ crystals such as the one used in our tests. The radiation (in red) is concentrated along the focusing direction and the image is produced at a distance equal to half the curvature radius of the crystal. Left: Experimental image of a concentrated 150 keV X-ray beam obtained at the LARIX Facility. The area of the crystal is of 30$\times$10 mm$^2$, while the area of the image is of 0.1$\times$10 mm$^2$, which means that the beam was concentrated in an area 30 times smaller than the size of the crystal itself.}
    \label{fig:concentration}
\end{figure}

Indeed, according to the dynamical theory of diffraction in bent perfect crystals, the presence of secondary curvature of the planes increases the reflectivity of the crystal above the limit of 50\%\cite{Authier98}. Intuitively, we can imagine that, since the planes are bent, the probability that a beam crossing the crystal undergoes a reflection is highly reduced, thus increasing the diffraction efficiency. 

For the purpose of building a broadband Laue lens with focusing capabilities and an optimized effective area, we aim to use, for our lens, bent perfect crystals of Silicon and Germanium with (111) diffracting planes. Tests done with mosaic Gallium Arsenide (220) crystals bent to a curvature radius of 40 m with a surface lapping technique\cite{Ferrari2013, Buffagni2015} allowed to experimentally verify that bent crystals are able to focus high energy X-rays\cite{Virgilli2016} (Fig. \ref{fig:concentration}). Further tests are being made with Ge(220) crystals and we plan to switch to perfect Si(111) and Ge(111), since the (111) planes should shows an increase of efficiency as described by Malgrange's theory of bent perfect crystal\cite{Malgrange2002}

\section{The TRILL project and its results}
\label{sec:trill}

\begin{figure}[t!]
    \centering
    \includegraphics[scale = 0.6]{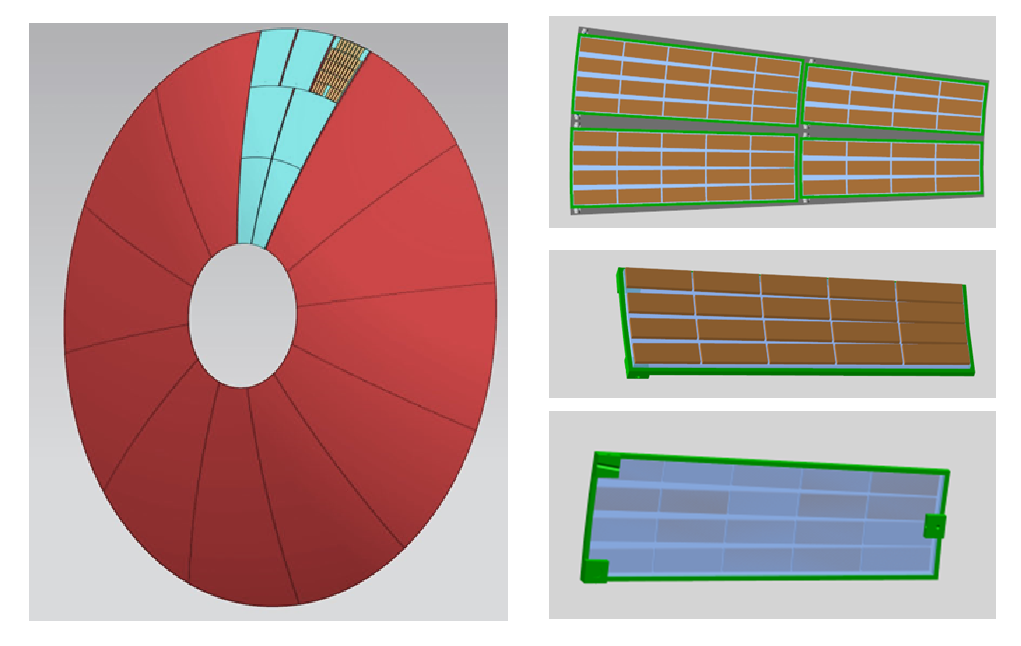}
    \caption{Left: Schematic CAD model of a full Laue lens. The lens is divided in spherical sectors, called petals (red), and each petal is divided in a series of modules (cyan). The crystals are fixed on each module. Top Right: CAD model of the four modules we are building joined together. Center Right: Front view of a single module with 20 crystals on top. Bottom Right: Rear view of a single module. The substrate on which the crystals are bounded is visible.}
    \label{fig:TRILL_CAD}
\end{figure}

The TRILL project (Technological Readiness Level Increase for Laue Lenses) was funded by the Italian Space Agency (ASI) through the National Institute for Astrophysics (INAF). Its aim is to increase the technological readiness level (TRL) of Laue lenses.
For our prototype of the Laue lens, we opted for a modular approach: a full lens is divided in spherical sectors ({\it petals}) and each sector is then divided into smaller {\it modules} which can contain up to 30 crystals each (Fig. \ref{fig:TRILL_CAD}). 

The crystals would be perfect crystal of Si and Ge with diffracting planes (111) and would be bent with the same curvature radius of $40$ m, which means that the focal length of the Laue lens for a source at infinite distance is $20$ m. 

Aim of the TRILL project is to build four modules and join them together in a first section of a petal. The desired PSF size of the whole lens must be of 30 arcsec, which translates in requiring that the Bragg's angle and the radial position of the crystal tiles are set within an accuracy better than 10 arcsec.

\label{sect:trill}

To reach those goals, the project is structured with the following main tasks: 
\begin{enumerate}
    \item To define a reliable and repeatable way to bend the crystals with no deterioration of their properties. The curvature radius of the bent Si and Ge crystals is required to be 40 m with an accuracy of $\pm~2$~m, and must be as uniform as possible to avoid distortion on the lens PSF.
    \item To find the best materials and the bonding method to build a module of Laue lens with the desired $<$10~arcsec orientation accuracy for the crystals, and to find a way to assemble together multiple modules. 
    \item To test the prototype with a suitable focal plane detector. {In order to reach an efficiency of the order of 80\% at 
    600 keV, a solid solid state, highly segmented, spectral-imager detector has been chosen. This choice will allow to fully exploit the focusing capabilities of the lens both for imaging, for spectroscopy and polarimetry.}
\end{enumerate}

In this section, we illustrate the status of advancement of each of those 3 tasks. 

\subsection{Crystals' manufacturing}
\begin{figure}[t!]
    \centering
    \includegraphics[scale = 0.5]{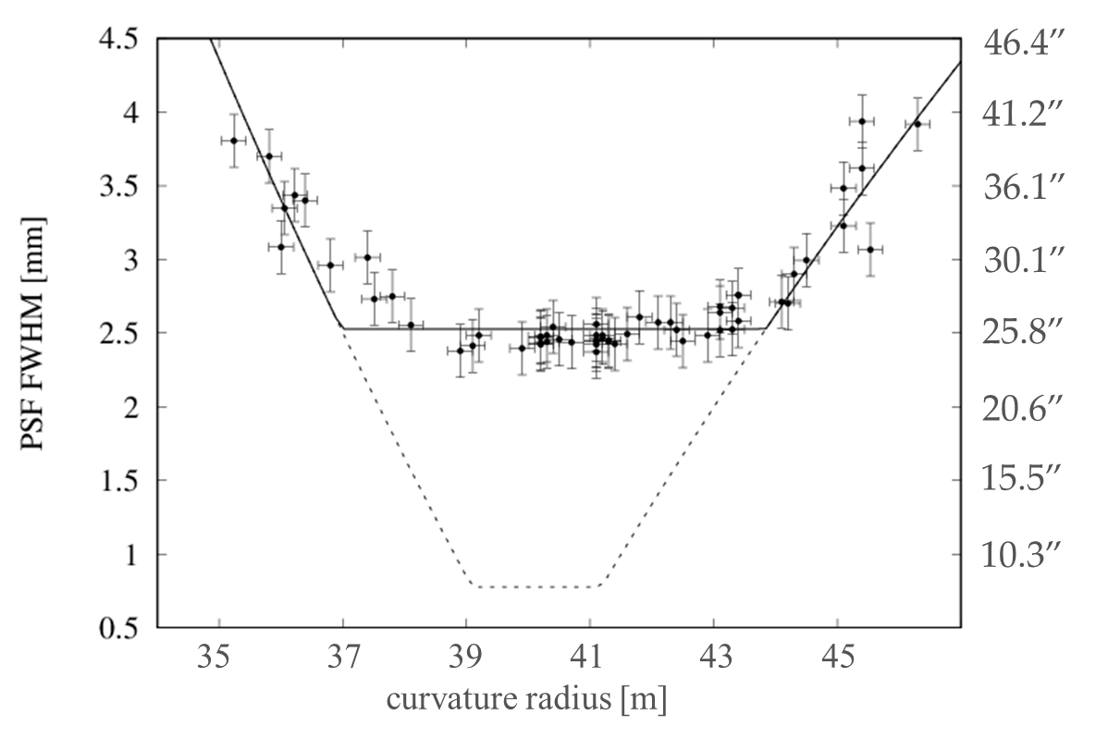}
    \caption{Black points: measured size of PSF of a sample of GaAs crystals versus their curvature radius. It can be immediately seen that between 38 m and 44 m of curvature radius, the deformation of the crystals' curvature radius from the optimal value has no effect on the size of the PSF. Black solid line: fit of the data points. Black dashed line: Theoretical PSF FWHM vs curvature radius curve for a Ge(111) crystal with quasi-mosaicity of 5 arcsec. The region in which the defocusing prevails on the radial deformation is reduced to about 2 m around the optimal value of 40 m.}
    \label{fig:pfs_vs_rad}
\end{figure}

The manufacturing of the bent perfect crystals of Silicon and Germanium is done at the CNR/IMEM institute in Parma (Italy). Dislocation-free wafers of Si and Ge of diameter of 4 inches are cut to obtain about 15 crystals of size 30 x 10 x 2 mm$\mathrm{{}^3}$. The crystals are cut in such a way that the chosen diffracting planes are (220). 
These planes are parallel to the 10 x 2 mm$\mathrm{{}^2}$ edge, with a {miscut} error of about 0.2 degrees.
Each tile is bent to the nominal curvature radius of  {40~m through a surface lapping method}. This technique consists in lapping one side of the crystal inducing a controlled damage which generates an internal strain able to bend the crystal in a self-standing way\cite{Ferrari2013, Buffagni2015}.  The first tests of this technique, done with GaAs(220) mosaic crystals, showed that  there is a range of values of curvature radii (see Fig.~\ref{fig:pfs_vs_rad}) for which the enlargement of the PSF of the crystal, induced by a not nominal curvature radius, is hidden by the defocusing effect due to the dispersion of the diffracting planes. This allows us to reduce the constraints on the distribution of the curvature radius of the crystals depending on the value of the angular dispersion of the planes.
{The smaller is the angular spread of a crystal, i.e. the mosaicity or the quasi-mosaicity, the more strict is the requirement on the 
curvature radius compared with the nominal one.}

For perfect Si(111) and Ge(111) crystals, which we aim to use for the definitive version of the lens, we expect that the bending of the crystals will induce a dispersion of the diffraction planes ({\it quasi-mosaicity}) of the crystals of $\sim$5 arcsec, which means that if the curvature radius of the crystals is within $\pm 2$ m from the nominal value of $40$ m, then the distortion of the PSF given by crystals with a slightly wrong curvature radius is negligible. {This means that the technique to bend the crystals that we are developing must guarantee that all the produced bent crystals have a curvature radius in the range $\mathrm{40 \pm 2}$~m.}

\begin{figure}[t!]
    \centering
    \includegraphics[scale = 0.6]{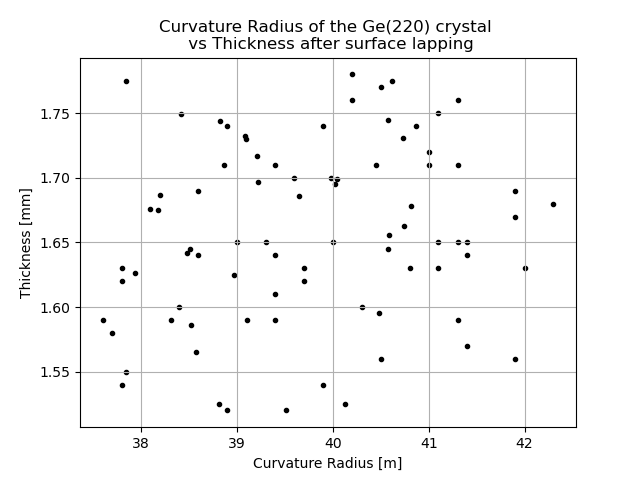}
    \caption{Curvature radius of the sample of 82 bent crystals of Ge(220) vs their thickness after the surface lapping procedure. Each point represents a crystal. All the curvature radii are within the limit $\pm$2 m from the nominal curvature radius of 40 m.}
    \label{fig:RvsT}
\end{figure}

In the context of the TRILL project, the surface lapping technique was tested with Ge(220) crystals. The average value of the curvature radii distribution is of 39.7 $\pm$ 0.2 m, with an estimate of the standard deviation of 1~m (Fig. \ref{fig:RvsT}), meaning that all the crystals are suitable for the 
{TRILL goals}. The curvatures of the crystals {were} measured at CNR/IMEM with a Phillips X'Pert PRO high resolution X-ray diffractometer, equipped with a Cu K$\mathrm{\alpha}$ source.

The lapping procedure shows two major downsides. The first is that it necessarily removes part of the material, therefore the thickness of the crystals significantly decreased with respect to the initial thickness.
The average value of the final thickness of the crystals after the lapping procedure is of (1.656 $\pm$ 0.002) mm (Fig. \ref{fig:RvsT}). The second downside of this technique is that, {to date, crystals with thickness higher than 2~mm are not bendable at the nominal 40~m radius. This represents an important drawback given that, especially for the higher energies, thicker crystals are required for optimizing the crystals diffraction efficiency}. Even with those limitations, the surface lapping technique results to be a very accurate, reliable and repeatable procedure to prepare crystals for a Laue lens.

\subsection{Module and petal assembly}
\begin{figure}[t!]
    \centering
    \includegraphics[scale = 0.8]{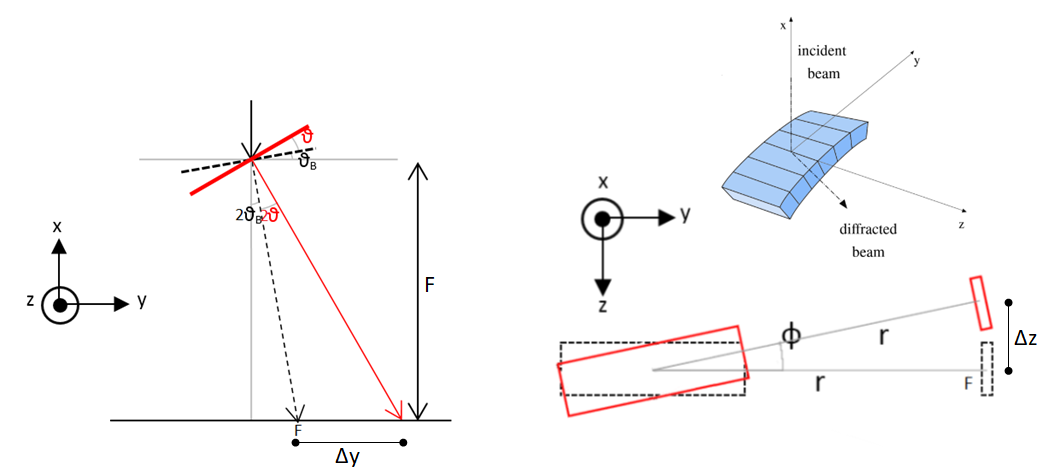}
    \caption{Schematics of the two main angles that need to be set to orient a crystal: the Bragg's angle (left) and the azimuth angle of the crystal on the ring (right). If a misalignment error on the angles is present (red), the image of the crystal will be shifted with respect to the expected position (black). In this drawing, $F$ is the focal length of the lens and $r$ is the distance from the centre of the lens at which the crystal is positioned, i.e. the radius of the ring on which the crystal is positioned. }
    \label{fig:angles}
\end{figure}

The second goal of the TRILL project is to define a method to build a Laue lens module by bonding the crystals on an adequate substrate with an accuracy sufficient to obtain a full lens PSF with 30~arcsec HPD. 
If the crystals are placed in concentric rings, two angles define the position of each tile: the Bragg's angle at which the crystals are set and the azimuth angle which indicates their position on the ring. The presence of a misalignment angle has the effect to shift the PSF from the desired focal point, meaning that for a full lens, the combined effect of the shifts of all the crystals' PSFs will result in a broadening of the PSF of the lens. 
It is then crucial to keep the misalignment errors as small as possible. The error on the Bragg's angle is the more critical, since the shift it induces is proportional to the focal length of the lens (Fig. \ref{fig:angles}). In this paper we will use the name $\theta$-misalignment to indicate the misalignment on the Bragg's angle and the $\phi$-misalignment to indicate the misalignment on the azimuth angle.

Ray-tracing simulations of a full Laue lens made of bent perfect crystals of Ge(111) established that it is necessary that all the crystals of a module are oriented  with an accuracy, on both angles, of  $\mathrm{<}$10 arcsec to obtain the desired PSF of 30 arcsec (FWHM) for the full lens, which results in a positional accuracy of the crystals of the order of a few microns. High precision bonding techniques need to be developed and applied to fulfill such a technologically challenging task.

We decided to bond the crystals to a suitable substrate by means of a low-shrinkage adhesive. The low-shrinkage requirement comes from the fact that it is crucial that the natural shrinkage of the glue during the curing process is as small as possible to reduce and control the misalignment on the crystal positions induced by the bonding procedure. Furthermore, to continuously test the positions of the crystals during the process, we use a hard X-ray horizontal beam, so the crystals must stay in vertical position during the bonding process. This means that the glue must be viscous enough to not drip down after its deposition.
\begin{figure}[t!]
    \centering
    \includegraphics{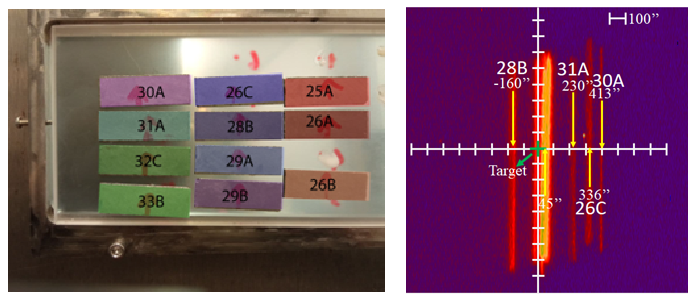}
    \caption{Left: The eleven Ge(220) crystals fixed on the quartz substrate. Right: Combined, real-time image of the 11 crystals. The 4 outlier crystal and their misalignment are highlighted.}
    \label{fig:ge_220_module}
\end{figure}

\begin{figure}[t!]
    \centering
    \includegraphics[scale = 0.7]{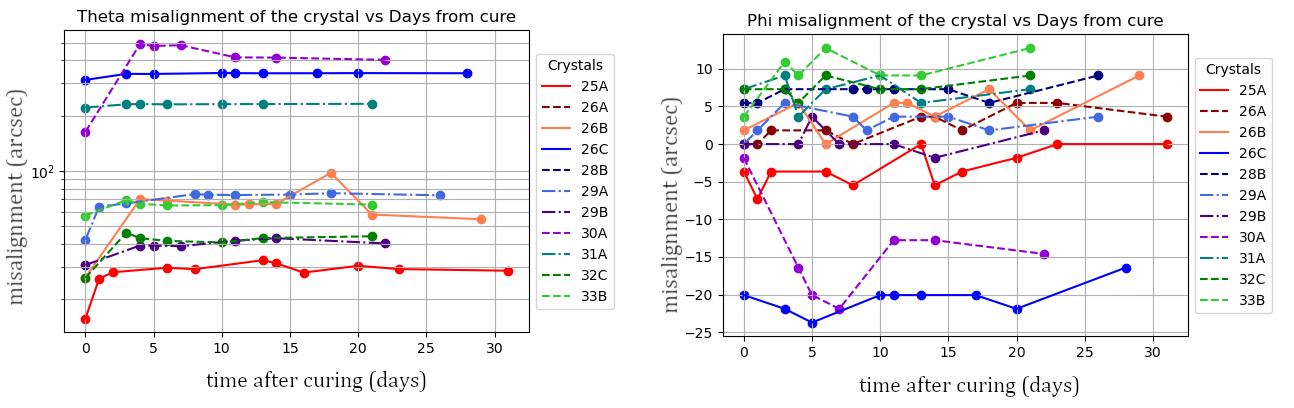}
    \caption{Left: $\theta$-Misalignment of the crystals vs time from bonding. Right:$\phi$-Misalignment of the crystals vs time from bonding.}
    \label{fig:misal_ge220}
\end{figure}

{The used adhesive is a single component UV-curable adhesive 
with a linear shrinkage of 0.03\% and a viscosity of 60,000 cP. The substrate was chosen in order to satisfy the following requirements:}
1) the hard X-rays must pass through the material without being heavily absorbed; 2) it should be transparent enough to UV-light to allow it to pass through the substrate to guarantee the complete cure of the adhesive; 3) it should have a low coefficient of thermal expansion (CTE). 
To satisfy all three requirements we chose a substrate of quartz 5~mm thick, since the quartz is transparent to UV, has an X-ray transmission sufficiently high ($\sim$84\% at 150 keV) and a low CTE ($\mathrm{5.5 \times 10^{-7}~K^{-1}}$). Other materials, such as Zerodur and Ultra Low Expansion Titanium Silicate glass, are under consideration.
Each substrate is shaped as an isosceles trapezium with bases of 68 mm and 56 mm long, respectively,  while the height is 183 mm. Given that the cross section of the crystal tiles is 30$\times$10 mm$^2$ and that the spacing between the crystals glued on the substrate will be of about 1 mm, each substrate can hold about 30 crystals. The surface of the substrate, on which the tiles are glued, was worked at the CNR/INO Institute in Arcetri (Florence, Italy), to impress a curvature radius of 40 m, which is the cylindrical curvature of the crystal tiles. To glue the tiles to the substrate, this is hold in vertical position though an INVAR steel frame.
The crystals are positioned on the glass substrate using a motorized hexapod with six degrees of freedom: three translations and three rotations. The hexapod has an accuracy of the order of 1~$\mu$m for translations and of 2$\times$10$^{-5}$~radians for rotations.

\begin{figure}[t!]
    \centering
    \includegraphics[scale = 1]{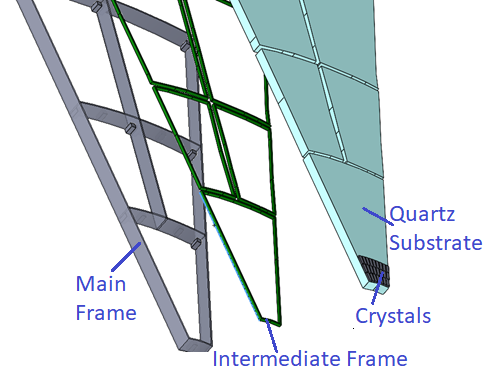}
    \caption{Exploded view of a Laue lens petal assembly. The crystals are bounded on a quartz substrate, then the substrate is fixed to a steel main frame with an intermediate frame of flexible material.}
    \label{fig:full_prot}
\end{figure}

To test the correct alignment of the crystals, we hit them with a collimated, polychromatic hard X-ray beam (beam cross section of 10 x 10 mm$\mathrm{^{2}}$) and collect the diffracted image with a Perkin Elmer hard X-ray imager (sensitive area 20 x 20 cm$\mathrm{{}^2}$ with spatial resolution of 200 $\mathrm{\mu}$m). The shift of the image from a target position, usually the center of the detector, allows us to evaluate how much the crystal is misaligned from the desired position.    

The bonding procedure is the following:
\begin{enumerate}
    \item the crystal is positioned on the hexapod and oriented to the proper Bragg's angle under the X-ray beam;
    \item a small drop of glue is deposited on the substrate with a glue dispenser;
    \item the crystal is placed in contact with the drop of glue. {We check that the crystal is still in the correct position. An angular correction can still be applied at this stage;}
    \item the glue is cured with UV light;
    \item the crystal is released from the hexapod and its final 
    position is evaluated immediately after the cure.
\end{enumerate}
To test the time stability of the bonding, we checked the position of the crystals each day. With this method we have built one module made of GaAs(220) crystals already available at our lab and one made of Ge(220) crystals. 
For the first prototype module, we tested only our capabilities to reach the desired accuracy on the Bragg's angle position and we found that the misalignment of the crystals after 50 days from bonding is within the interval +15/-10 arcsec. 
For the second prototype, we tested our capabilities to correctly set both the Bragg's angle and the azimuth angle of the crystals. Eleven crystals of Ge(220) were fixed on the substrate following the procedure described above and the first real-time image of the PSF of all the crystals combined together was obtained (Fig.\ref{fig:ge_220_module}).
For this second test we found that, after about 25 days from the cure, the average value of the $\theta$-misalignment is of 101~arcsec, with a standard deviation of 161~arcsec (see left panel of Fig.~\ref{fig:misal_ge220}). 
Four crystals (28B, 26C, 30A, 31A) results to be strongly misaligned. We suspect that this may be due to a deformation of the region of the substrate on which they are glued.  The average value of the $\phi$-misalignment is found to be 2~arcsec, with a standard deviation of 9~arcsec. The $\phi$-misalignment versus time from the cure is shown in the right panel of Fig.~\ref{fig:misal_ge220}.

These results state that the required accuracy on the $\phi$-misalignment has been achieved, while we are still one order of magnitude far from reaching the same results on the $\theta$-misalignment.
One positive aspect is that, once cured, the adhesive guarantees a good stability of the assembly on long time scale. 
We argue that the $\theta$-misalignment is due to the technical difficulty of distributing the layer of adhesive uniformly and with a controlled thickness. The next tests will be dedicated to control the  distribution of the adhesive with an accuracy of a tenth of microns.

Finally, the development of a way to assemble the modules together is currently in a study phase. At the moment, we plan to join the modules together by fixing them on an intermediate frame. The frame will be oriented with respect to a main rigid frame through high precision micrometric screws (Fig. \ref{fig:full_prot}).

\subsection{The focal plane detector}
\begin{figure}[t!]
    \centering
    \includegraphics[scale = 0.75]{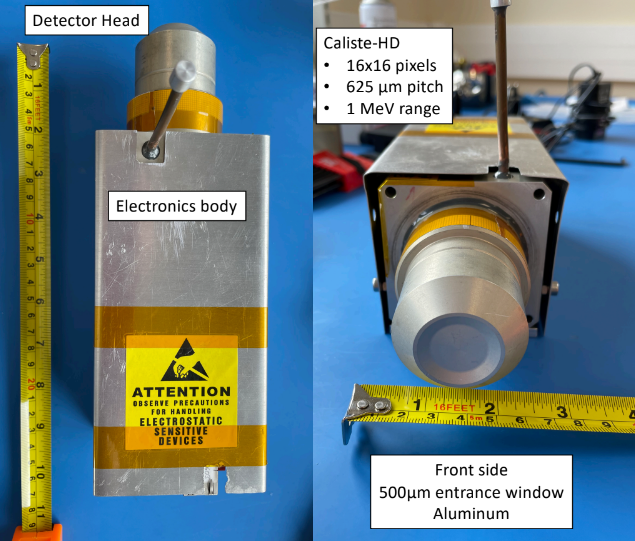}
    \caption{The WIX-HD setup, which contains the Caliste-HD detector and its mechanical structures, enclosed in a tight thermal controlled atmosphere. The box contains the read-out electronics. }
    \label{fig:caliste}
\end{figure}

To exploit at best the focusing capabilities of a Laue lens, the best focal plane detector is a highly-segmented spectral-imager detector with 3-D spatial resolution. In this way it is possible to perform spectroscopic measurements and imaging of a source at the same time. Both the segmentation and the 3-D resolution, together with a very good spectral resolution, allow us to use the detector also as a Compton polarimeter. Solid state, segmented detectors based on Cadmium Telluride (CdTe) or Cadmiun Zinc Telluried (CZT) are the candidate solid state materials for our scope, since their high atomic number makes them very efficient for high energy photons. Furthermore, they can be used at room temperature or with a minimal cooling made by a Peltier cell, thus they are easy to handle and with a low power consumption and weight, making them great candidate for a space mission detector.

Different spectral-imager detectors were taken into consideration for a combined use with Laue lenses. In the context of the TRILL project, we decided that the best detector for testing our Laue lens prototype is the Caliste-HD device\cite{Meuris2011, Maier2018}, developed at CEA (Fig. \ref{fig:caliste}). Caliste-HD is made by an Al/CdTe/Pt pixel detector (leakage current $<$10 pA) and is fully compliant with space operations (radiation resistant, and resistant to thermal and mechanical instabilities). The sensitive area of the detector is 1 x 1 cm$\mathrm{{}^2}$, with a crystal thickness of 1 mm and a pixel size of 0.625 x 0.625 mm$\mathrm{{}^2}$. The pass-band of the detector is from 2 keV to 1 MeV and it has an energy resolution of 1.1\% at 60 keV.

At the moment, the spectroscopic and imaging capabilities of the Caliste-HD detector are being studied at INAF-OAS Bologna (Italy). In order to evaluate the spectral-imaging capability of the detector, it will be used in combination with the Ge(220) Laue lens module.

\section{The future: space missions based on Laue lenses}
\label{sec:future}
\begin{figure}[t!]
    \centering
    \includegraphics[scale = 0.5]{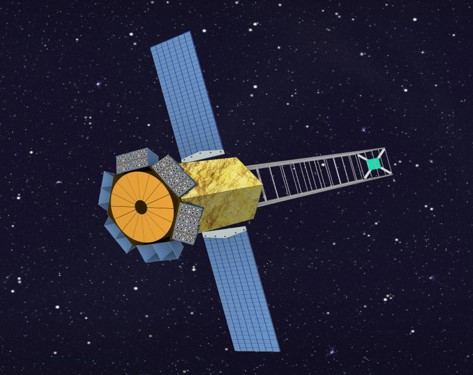}
    \caption{Artistic representation of ASTENA's in flight configuration. The WFM-I/S units surrounds the body of the spacecraft, on which the Laue lens (orange) will be put on top. The focal plane detector (cyan) will be kept at a distance of 20 m from the lens by means of an extensible mast.}
    \label{fig:ASTENA}
\end{figure}

Our studies on Laue lenses, including the TRILL project, are aimed to develop the technology for building the Narrow Field Telescope (NFT) on-board ASTENA, the Advanced Surveyor for Transient Events and Nuclear Astrophysics, a concept mission that we proposed to achieve the goals discussed in two white papers we submitted to ESA for its long term program "Voyage 2050" \cite{Frontera2021,Guidorzi2021} (Fig. \ref{fig:ASTENA}). ASTENA's payload consists of two main instruments: the Wide Field Monitor - Imager/Spectrometer (WFM-I/S) and the Narrow Field Telescope (NFT). The WFM-I/S consists of twelve Position Sensitive Detector (PSD) units surmounted by a double-scale coded mask, one of which (for low energies) is one-dimensional. The PSD is based on the same principles of the X/Gamma-ray Imaging Spectrometer (XGIS) proposed for the THESEUS space mission \cite{Campana2018}.
The NFT will be a revolutionary hard X/soft Gamma-ray focusing telescope working in the energy band 50-700 keV based on the technology of Laue lenses. At the moment, a possible pathfinder of the ASTENA mission concept, with only the NFT, is also under study, for a satellite or balloon experiment. The NFT is a Laue lens of 3 m in diameter with a 20 m focal length. The lens will be made of about 20 thousand crystal tiles of perfect Si(111) and Ge(111) of cross section 30$\times$10 mm$^2$ and thickness optimized for the diffraction efficiency vs. energy. The crystals will be radially bent with a curvature radius of 40 m. In the context of the TRILL project, we are developing and testing some possible techniques that will be used for manufacturing the crystals and build the whole lens.  The focal plane detector will be a pixelated CdZnTe spectral-imager detector, with a size of 8$\times$8$\times$8 cm$^3$, a pitch of 300 $\mu$m and an efficiency $>$80\% in the whole energy band of NFT.  In this configuration, the NFT is expected to achieve an unprecedented angular resolution in the sub-MeV energy range of 30 arcsec and a point source localization accuracy $<$10 arcsec, with a Field of View of 4 arcmin. The NFT will bring a leap in sensitivity of two order of magnitude with respect to the best current instruments operating in the same energy bands, opening a new range of possibilities for high energy astronomical observations.

\acknowledgments
 
This work is partly supported by the AHEAD-2020 Project grant agreement 871158
of the European Union’s Horizon 2020 Programme and by the ASI-INAF agreement
no. 2017-14-H.O ”Studies for future scientific missions”.

\bibliography{report}

\begin{thebibliography}{10}

\bibitem{Zachariasen}
Zachariasen, W.~H.,  [{\em Theory of X-ray Diffraction in
  Crystals}{\nolinebreak\hspace{0.1em}]}, Wiley (1945).

\bibitem{frontera2011}
{Frontera}, F. and {von Ballmoos}, P., ``{Laue Gamma-Ray Lenses for Space
  Astrophysics: Status and Prospects},'' {\em X-Ray Optics and
  Instrumentation}~{\bf 2010},  215375 (Jan. 2010).

\bibitem{Guidorzi2021}
Guidorzi, C., Frontera, F., Ghirlanda, G., Stratta, G., Mundell, C.~G.,
  Virgilli, E., Rosati, P., Caroli, E., Amati, L., Pian, E., Kobayashi, S.,
  Ghisellini, G., Fryer, C., Valle, M.~D., Margutti, R., Marongiu, M., Martone,
  R., Campana, R., Fuschino, F., Labanti, C., Orlandini, M., Stephen, J.~B.,
  Brandt, S., Silva, R. C.~d., Laurent, P., Mochkovitch, R., Bozzo, E., Ciolfi,
  R., Burderi, L., and Di~Salvo, T., ``A deep study of the high--energy
  transient sky,'' {\em Experimental Astronomy}~{\bf 51},  1203--1223 (Jun
  2021).

\bibitem{Frontera2021}
Frontera, F., Virgilli, E., Guidorzi, C., Rosati, P., Diehl, R., Siegert, T.,
  Fryer, C., Amati, L., Auricchio, N., Campana, R., Caroli, E., Fuschino, F.,
  Labanti, C., Orlandini, M., Pian, E., Stephen, J.~B., Del~Sordo, S.,
  Budtz-Jorgensen, C., Kuvvetli, I., Brandt, S., da~Silva, R. M.~C., Laurent,
  P., Bozzo, E., Mazzali, P., and Valle, M.~D., ``Understanding the origin of
  the positron annihilation line and the physics of supernova explosions,''
  {\em Experimental Astronomy}~{\bf 51},  1175--1202 (Jun 2021).

\bibitem{frontera2008}
Frontera, F., Loffredo, G., Pisa, A., Nobili, F., Carassiti, V., Evangelisti,
  F., Landi, L., Squerzanti, S., Caroli, E., Stephen, J.~B., Andersen, K.~H.,
  Courtois, P., Auricchio, N., Milani, L., and Negri, B., ``{Focusing of
  gamma-rays with Laue lenses: first results},'' in [{\em Space Telescopes and
  Instrumentation 2008: Ultraviolet to Gamma Ray}{\nolinebreak\hspace{0.1em}]},
   Turner, M. J.~L. and Flanagan, K.~A., eds.,  {\bf 7011},  568 -- 575,
  International Society for Optics and Photonics, SPIE (2008).

\bibitem{Authier98}
Authier, A. and Malgrange, C., ``{Diffraction Physics},'' {\em Acta
  Crystallographica Section A}~{\bf 54},  806--819 (Nov 1998).

\bibitem{Camattari2015}
Camattari, R., Guidi, V., Bellucci, V., and Mazzolari, A., ``The `quasi-mosaic'
  effect in crystals and its applications in modern physics,'' {\em Journal of
  Applied Crystallography}~{\bf 48} (08 2015).

\bibitem{virgilli2014}
Virgilli, E., Frontera, F., Valsan, V., Liccardo, V., Carassiti, V.,
  Squerzanti, S., Statera, M., Parise, M., Chiozzi, S., Evangelisti, F.,
  Caroli, E., Stephen, J., Auricchio, N., Silvestri, S., Basili, A., Cassese,
  F., Recanatesi, L., Guidi, V., Bellucci, V., Camattari, R., Ferrari, C.,
  Zappettini, A., Buffagni, E., Bonnini, E., Pecora, M., Mottini, S., and
  Negri, B., ``The {LAUE} project and its main results,'' (2014).

\bibitem{Ferrari2013}
{Ferrari}, C., {Buffagni}, E., {Bonnini}, E., and {Zappettini}, A., ``{X-ray
  diffraction efficiency of bent GaAs mosaic crystals for the LAUE project},''
  in [{\em Optics for EUV, X-Ray, and Gamma-Ray Astronomy
  VI}{\nolinebreak\hspace{0.1em}]},  {O'Dell}, S.~L. and {Pareschi}, G., eds.,
  {\em Society of Photo-Optical Instrumentation Engineers (SPIE) Conference
  Series} {\bf 8861} (2013).

\bibitem{Buffagni2015}
{Buffagni}, E., {Bonnini}, E., {Ferrari}, C., {Virgilli}, E., and {Frontera},
  F., ``{X-ray characterization of curved crystals for hard x-ray astronomy},''
  in [{\em EUV and X-ray Optics: Synergy between Laboratory and Space
  IV}{\nolinebreak\hspace{0.1em}]},  {Hudec}, R. and {Pina}, L., eds., {\em
  Society of Photo-Optical Instrumentation Engineers (SPIE) Conference Series}
  {\bf 9510} (2015).

\bibitem{Virgilli2016}
{Virgilli}, E., {Frontera}, F., {Rosati}, P., {Bonnini}, E., {Buffagni}, E.,
  {Ferrari}, C., {Stephen}, J.~B., {Caroli}, E., {Auricchio}, N., {Basili}, A.,
  and {Silvestri}, S., ``{Focusing effect of bent GaAs crystals for
  {\ensuremath{\gamma}}-ray Laue lenses: Monte Carlo and experimental
  results},'' {\em Experimental Astronomy}~{\bf 41},  307--326 (Feb. 2016).

\bibitem{Malgrange2002}
Malgrange, C., ``X-ray propagation in distorted crystals: From dynamical to
  kinematical theory,'' {\em Crystal Research and Technology}~{\bf 37}(7),
  654--662 (2002).

\bibitem{Meuris2011}
Meuris, A., Limousin, O., Gevin, O., Lugiez, F., Le~Mer, I., Pinsard, F.,
  Donati, M., Blondel, C., Michalowska, A., Delagnes, E., Vassal, M.-C., and
  Soufflet, F., ``Caliste {HD}: {A} new fine pitch {C}d({Z}n){T}e imaging
  spectrometer from 2 kev up to 1 {M}ev,'' in [{\em 2011 IEEE Nuclear Science
  Symposium Conference Record}{\nolinebreak\hspace{0.1em}]},   4485--4488
  (2011).

\bibitem{Maier2018}
Maier, D., Blondel, C., Delisle, C., Limousin, O., Martignac, J., Meuris, A.,
  Visticot, F., Daniel, G., Bausson, P.-A., Gevin, O., Amoyal, G., Carrel, F.,
  Schoepff, V., Mahé, C., Soufflet, F., and Vassal, M.-C., ``Second generation
  of portable gamma camera based on {C}aliste {C}d{T}e hybrid technology,''
  {\em Nuclear Instruments and Methods in Physics Research Section A:
  Accelerators, Spectrometers, Detectors and Associated Equipment}~{\bf 912},
  338--342 (2018).
\newblock New Developments In Photodetection 2017.

\bibitem{Campana2018}
{Campana}, R., {Fuschino}, F., {Labanti}, C., {Amati}, L., {Mereghetti}, S.,
  {Fiorini}, M., {Frontera}, F., {Baldazzi}, G., {Bellutti}, P., {Borghi}, G.,
  {Elmi}, I., {Evangelista}, Y., {Feroci}, M., {Ficorella}, F., {Orlandini},
  M., {Picciotto}, A., {Marisaldi}, M., {Rachevski}, A., {Uslenghi}, M.,
  {Vacchi}, A., {Zampa}, G., {Zampa}, N., and {Zorzi}, N., ``{The X-Gamma
  Imaging Spectrometer (XGIS) onboard THESEUS},'' {\em Mem. Societa Astronomica
  Italiana}~{\bf 89},  137 (Jan. 2018).

\end{thebibliography}
\bibliographystyle{spiebib} 

\end{document}